\documentclass[10pt,conference]{IEEEtran}
\ifCLASSINFOpdf
\else
\fi
\usepackage[latin9]{inputenc}
\usepackage{amsmath}
\usepackage{array}
\usepackage{amssymb}
\usepackage{mathrsfs}
\usepackage{amsbsy}
\usepackage{listings}
\usepackage{citesort}
\usepackage{graphicx}
\usepackage{multirow}
\usepackage{mathtools}
\usepackage{url}
\usepackage{color}
\usepackage{soul}
\usepackage{booktabs}
\usepackage[table,xcdraw]{xcolor}
\usepackage{booktabs,caption,fixltx2e}
\usepackage[flushleft]{threeparttable}
\usepackage{flushend}
\usepackage{listings}
\usepackage[latin9]{inputenc}
\usepackage{stackengine}

\newlength\llength
\definecolor{mygreen}{rgb}{0,0.6,0}
\definecolor{mygray}{rgb}{0.5,0.5,0.5}
\definecolor{mymauve}{rgb}{0.58,0,0.82}

\global\long\def\Sb{\boldsymbol{s}}

\global\long\def\xb{\boldsymbol{x}}

\global\long\def\bb{\boldsymbol{b}}

\global\long\def\wb{\boldsymbol{w}}
\global\long\def\xb{\boldsymbol{x}}
\global\long\def\yb{\boldsymbol{y}}

\global\long\def\bb{\boldsymbol{b}}

\global\long\def\sigmab{\boldsymbol{\sigma}}

\begin{document}

\title{Automatic feature learning for vulnerability prediction}

\author{\IEEEauthorblockN{Hoa Khanh Dam\IEEEauthorrefmark{1},
		  Truyen Tran\IEEEauthorrefmark{2},
            Trang Pham\IEEEauthorrefmark{2},
            Shien Wee Ng\IEEEauthorrefmark{1},
            John Grundy\IEEEauthorrefmark{2} and
		  Aditya Ghose\IEEEauthorrefmark{1}}
		
\IEEEauthorblockA{\IEEEauthorrefmark{1}
	University of Wollongong, Australia\\
	Email: \{hoa,swn881,aditya\}@uow.edu.au}
\IEEEauthorblockA{\IEEEauthorrefmark{2}
	Deakin University, Australia\\
	Email: \{truyen.tran,phtra,j.grundy\}@deakin.edu.au}
}

\maketitle

\begin{abstract}
Code flaws or vulnerabilities are prevalent in software systems and can potentially cause a variety of problems including deadlock, information loss, or system failure. A variety of approaches have been developed to try and detect the most likely locations of such code vulnerabilities in large code bases. Most of them rely on manually designing features (e.g. complexity metrics or frequencies of code tokens) that represent the characteristics of the code. However, all suffer from challenges in sufficiently capturing both semantic and syntactic representation of source code, an important capability for building accurate prediction models. In this paper, we describe a new approach, built upon the powerful deep learning Long Short Term Memory model, to automatically learn both semantic and syntactic features in code. Our evaluation on 18 Android applications demonstrates that the prediction power obtained from our learned features is equal or even superior to what is achieved by state of the art vulnerability prediction models: 3\%--58\% improvement for within-project prediction and 85\% for cross-project prediction.



\end{abstract}

\section{Introduction}


A software vulnerability -- a security flaw, glitch, or weakness found in software systems -- can potentially cause significant damages to businesses and people's lives, especially with the increasing reliance on software in all areas of our society. For instance, the Heartbleed vulnerability in OpenSSL exposed in 2014 has affected billions of Internet users \cite{Heartbleed}. Cyberattacks are constant threats to businesses, governments and consumers. The rate and cost of a cyber breach is increasing rapidly with annual cost to the global economy from cybercrime being estimated at \$400 billion \cite{McAfee}. In 2017, it is estimated that the global security market is worth \$120 billion \cite{CybersecurityVentures}. Central to security protection is the ability to detect and mitigate software vulnerabilities early, especially before software release to effectively prevent attackers from exploit them.



Software has significantly increased in both size and complexity. Identifying security vulnerabilities in software code is highly difficult since they are rare compared to other types of software defects. For example, the infamous Heartbleed vulnerability was caused only by two missing lines of code \cite{Heartbleed2014}. Finding software vulnerabilities is commonly referred to as ``searching for a needle in a haystack'' \cite{Zimmermann:2010:SNH}. Static analysis tools have been routinely used as part of the security testing process but they commonly generate a large number of false positives \cite{Austin:2011:OTE,Ceccato:2016:SAP}. Dynamic analysis tools rely on detailed monitoring of run-time properties including log files and memory, and require a wide range of representative test cases to exercise the application. Hence, standard practice still relies heavily on domain knowledge to identify the most vulnerable part of a software system for intensive security inspection.

Software engineers can be supported by automated tools that explore remaining parts of the code base more likely to contain vulnerabilities and raise an alert on these. Such predictive models and tools can help prioritize  effort and optimize inspection and testing costs. They aim to increase the likelihood of finding vulnerabilities and reduce the time required by software engineers to discover vulnerabilities. In addition, a predictive capability that identifies vulnerable components early in the software lifecycle is a significant achievement since the cost of finding and fixing errors increases dramatically as the software lifecycle progresses.


A common approach to build vulnerability prediction models is by using machine learning techniques. A number of features representing software code are selected for use as predictors for vulnerability. The most commonly used features in previous work (e.g. \cite{Shin:2011:ECC}) are software metrics (e.g. size of code, number of dependencies, and cyclomatic complexity), code churn metrics (e.g. the number of code lines changed), and developer activity. Those features cannot however distinguish code regions of different semantics. In many cases, two pieces of code may have the same complexity metrics but they behave differently and  and thus have different likelihood of vulnerability to attack. Furthermore, the choice of which features are selected as predictors is \emph{manually} chosen by knowledgeable domain experts, and may thus carry outdated  experience and underlying biases. In addition, in many situations  handcrafted features normally do not generalize well:  features that work well in a  certain  software  project  may  not  perform  well  in  other projects \cite{Zimmermann:2009:CDP}.

An emerging approach is treating software code as a form of text and leveraging Natural Language Processing (NLP) techniques to automatically extract features. Previous work (e.g. \cite{ScandariatoWHJ14}) has used Bag-of-Words (BoW) to represent a source code file as a collection of code tokens associated with frequencies. The terms are the features which are used as the predictors for their vulnerability prediction model. Their set of features are thus not fixed or pre-determined (as seen in the software metric model), but rather depend on the vocabulary used by developers. However, the BoW approach has two major weaknesses. Firstly,  it ignores the semantics of code tokens, e.g. fails to recognize the semantic relations between ``for'' and ``while''. Secondly, a bag of code tokens does not necessarily capture the semantic structure of code, especially its sequential nature.




The recent advances of deep learning models \cite{lecun2015deep} in machine learning offer a powerful alternative to software metrics and BoW in representing software code. One of the most widely-used deep learning models is Long Short-Term Memory (LSTM) \cite{hochreiter1997long}, a special kind of recurrent neural networks that is highly effective in learning long-term dependencies in sequential data such as text and speech. LSTMs have demonstrated ground-breaking performance in many applications such as machine translation, video analysis, and speed recognition  \cite{lecun2015deep} .


This paper presents a novel deep learning-based approach to \emph{automatically learn features} for predicting vulnerabilities in software code. We leverage LSTM to capture the long context relationships in source code where dependent code elements are scattered far apart. For example, pairs of code tokens that are required to appear together due to programming language specification (e.g. $try$ and $catch$ in Java) or due to API usage specification (e.g. $lock()$ and $unlock()$), but do not immediately follow each other. The learned features sufficiently represent both the semantics of code tokens (\emph{semantic features}) and the sequential structure of source code (\emph{syntactic features}). Our automatic feature learning approach eliminates the need for manual feature engineering which occupies most of the effort in traditional approaches. Results from our experiments on 18 Java applications for the Android OS platform from a public dataset \cite{ScandariatoWHJ14} demonstrate that our approach is highly effective in predicting vulnerabilities in code.

The outline of this paper is as follows. In the next section, we provide a motivation example. Section III provides a brief background on vulnerability prediction and the neural networks used in our model. We then present our approach in Section IV, its implementation in Section V, report the experiments to evaluate it in Section VI, and discuss the threats to validity in Section VII. In Section VIII, we discuss related work before summarizing the contributions of the paper and outlines future work in Section IX.

\section{Motivation}

Figure \ref{fig:example} shows two code listings in Java which was adapted from \cite{CERT-Java}. Both pieces of code aim to avoid data corruption in multi-threaded Java programs by protecting shared data from concurrent modifications and accesses (e.g. file $f$ this example). They do so by using a reentrant mutual exclusion lock $l$  to enforce exclusive access to the file $f$. Here, a thread executing this code means to acquire the lock before reading file $f$, and then release the lock when it is done with the file to allow other threads to access the file.


\setbox0=\hbox{%
\begin{minipage}{1.5in}
\begin{lstlisting}[caption=File1.java]
try {
  l.lock()
  readFile(f);
  l.unlock();
}
catch (Exception e) {
  // Do something
}
finally {
  closeFile(f);
}
\end{lstlisting}
\end{minipage}
}
\savestack{\listingA}{\box0}

\setbox0=\hbox{%
\begin{minipage}{1.5in}
\begin{lstlisting}[caption=File2.java]
l.lock()
try {
  readFile(f);
}
catch (Exception e) {
  // Do something
}
finally {
  l.unlock();
  closeFile(f);
}
\end{lstlisting}
\end{minipage}
}
\savestack{\listingB}{\box0}

\begin{figure}[ht]
    \begin{tabular}{|c|c}
    \listingA &
    \listingB \\
    \end{tabular}
    \caption{A motivating example}
	\label{fig:example}
\end{figure}

The use of such locking can however result in deadlocks. Listing 1 in Figure \ref{fig:example} demonstrates an example of deadlock vulnerabilities. While it reads file $f$, an exception (e.g. file not found) may occur and control transfers to the $catch$ block. Hence, the call to $unlock()$ never gets executed, and thus it fails to release the lock. An unreleased lock in a thread will prevent other threads from acquiring the same lock, leading to a deadlock situation. Deadlock is a serious vulnerability, which can be exploited by attackers to organise Denial of Service (DoS) attacks. This type of attack can slow or prevent legitimate users from accessing a software system.

Listing 2 in Figure \ref{fig:example} rectifies this vulnerability. It fixes the problem of the lock not being released by calling $unlock()$ in the $finally$ block. Hence, it guarantees that the lock is released regardless of whether or not an exception occurs. In addition, the code ensures that the lock is held when the $finally$ block executes by acquiring the lock (calling $lock()$) immediately before the $try$ block.

The two code listings are identical with respect to both software metric and Bag-of-Words measures used by most current predictive and machine learning approaches. The number of code lines, the number of conditions, variables, and branches are the same in both listings. The code tokens and their frequencies are also identical in both pieces of code. Hence, the two code listings are indistinguishable if either software metrics or BoW are used as features. Existing work which relies on those features would fail to recognize that the left-hand side listing contains a vulnerability while the right-hand side does not.

\section{Background}

\subsection{Vulnerability prediction}\label{sect:vuln-prediction}

Vulnerability prediction typically involves determining whether a source code file is likely to be vulnerable. The goal is to alert software engineers with parts of the code base that deserve particular attention, rather than pinpointing exactly the code line where a vulnerability is resided. Hence, we also choose to work at the level of files since this is also the scope of existing work (e.g. \cite{ScandariatoWHJ14,Shin:2011:ECC}) which we would like to compare our approach against.

Determining if a source file is vulnerable can be considered as a function $vuln(x)$ which takes as input a file $x$ and returns a boolean value: $true$ indicates that the file is vulnerable, while $false$ indicates that the file is clean. 
Machine learning techniques have been widely used to learn function $vuln(x)$. To make it mathematically and computationally convenient for machine learning algorithms, file $x$ needs to be represented as a n-dimensional vector where each dimension represents a feature (or predictor). 






\subsection{Recurrent Neural Network and Long Short Term Memory}

\begin{figure}[ht]
	\centering
	\includegraphics[width=0.9\linewidth]{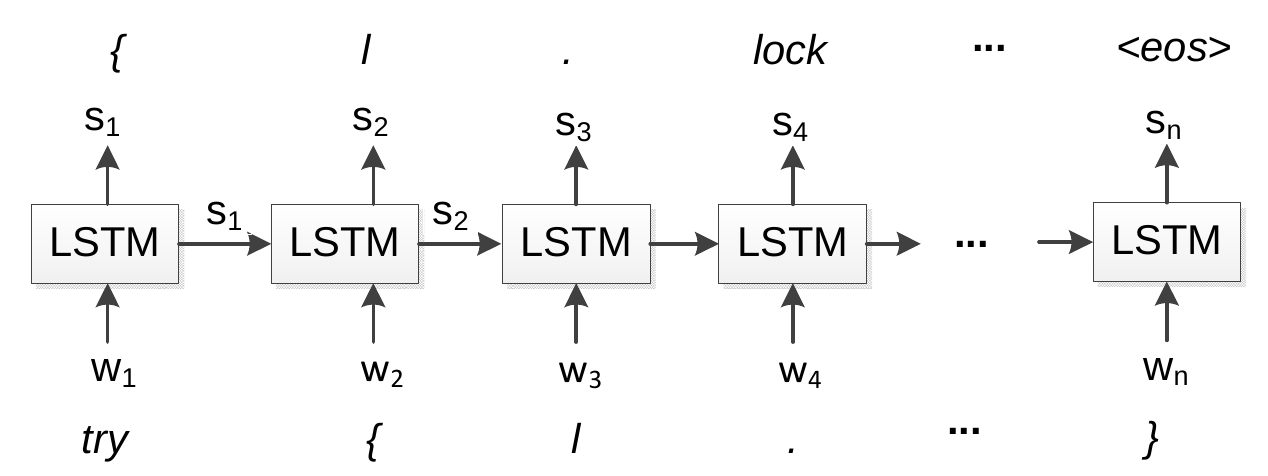}
	\caption{A recurrent neural network}
	\label{fig:RNN-LSTM}
\end{figure}

A Recurrent Neural Network (RNN) \cite{hochreiter2001gradient} is a single-hidden-layer neural network repeated multiple times. While a feedforward neural network maps an input vector into an output vector, an RNN maps \textbf{a sequence into a sequence} (see Figure \ref{fig:RNN-LSTM}). Let $\wb_{1},...,\wb_{n}$ be the input sequence (e.g. code tokens) and $\yb_{1},...,\yb_{n}$ be the sequence of corresponding labels (e.g. the next code tokens). At each step $t$, a standard RNN model reads the input $\wb_{t}$ and the previous output state $\Sb_{t-1}$ to compute the output state $\Sb_{t}$ as follows.

\begin{equation}\label{eq:hidden-state}
\Sb_{t}=\sigma\left(\bb+W_{tran}\Sb_{t-1}+W_{in}\wb_{t}\right)
\end{equation}
where $\sigma$ is a nonlinear element-wise transform function, and $\bb$, $W_{tran}$ and $W_{in}$
are referred to as \emph{model parameters}. The output state is used to predict the output (e.g. the next code token based on the previous ones) at each step $t$.


An RNN shares the same parameters across all steps since the same task is performed at each step, just with different inputs. Hence, using an RNN significantly reduces the total number of model parameters which we need to learn. The RNN model is trained using many input sequences with known true output sequences. The errors between the true outputs and the predicted outputs are passed backwards through the network during training to adjust the model parameters such that the errors are minimized.

\begin{figure}[ht]
	\centering
	\includegraphics[width=0.5\linewidth]{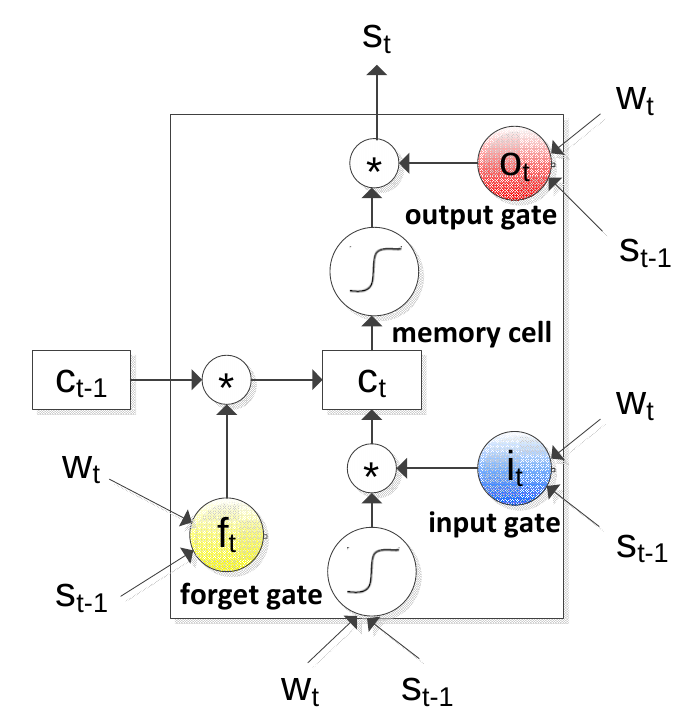}
	\caption{The internal structure of an LSTM unit}
	\label{fig:LSTMunit}
\end{figure}

A Long Short-Term Memory (LSTM) \cite{hochreiter1997long,gers2000learning} architecture is a special variant of RNN, which is capable of learning long-term dependencies. Due to space limitations, we briefly describe LSTM here and refer the readers to the seminal paper \cite{hochreiter1997long} for more details. It has a \emph{memory cell} $c_t$ which stores accumulated memory of the context. The amount of information flowing through the memory cell is controlled by three gates (an \emph{input gate} $i_t$, a \emph{forget gate} $f_t$, and an \emph{output gate} $o_t$), each of which returns a value between 0 (i.e. complete blockage) and 1 (full passing through).
All those gates are \emph{learnable}, i.e. being trained with the whole code corpus. It is important to note here that LSTM computes the output state based on not just only the current input $w_t$ and the previous ouput state $h_{t-1}$ (as done in standard RNNs) but also the current memory cell state $c_t$, which is \emph{linear} with the previous memory $c_{t-1}$. This is the key feature allowing an LSTM model to learn long-term dependencies.

\section{Approach}

\subsection{Overview}

\begin{figure}[ht]
	\centering
	\includegraphics[width=\linewidth]{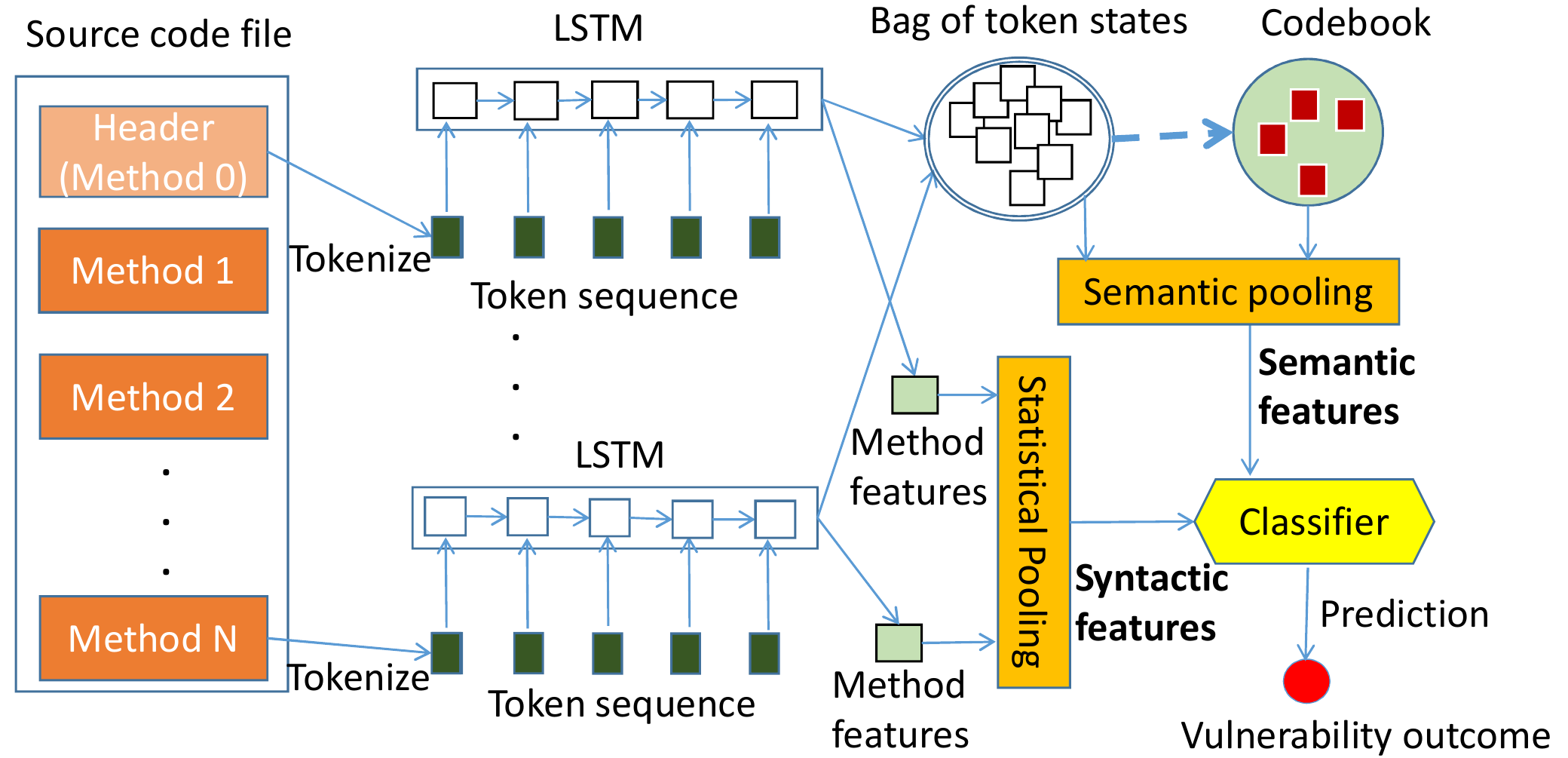}
	\caption{Overview of our approach for automatic feature learning for vulnerability prediction based on LSTM. The codebook is constructed from all bags of token states in all projects, and the process is detailed in Fig.~\ref{fig:codebook}}
	\label{fig:LSTM-rep}
\end{figure}

Our process of automatically learning and extracts both syntactic and semantic features goes through multiple steps (see Figure \ref{fig:LSTM-rep}).  These features used to build a classifier for vulnerability prediction. We consider each Java source file as consisting of a header (which contains a declaration of class variables) and a set of methods. We treat a header as a special method (method 0). We parse the code within each method into a \emph{sequence of code tokens}, which is fed into into a Long Short-Term Memory (LSTM) system to learn a vector representation of the method (i.e. \emph{method features}). This important step transforms a variable-size sequence of code tokens into a fixed-size feature vector in a multi-dimensional space. In addition, for each input code token, the trained LSTM system also gives us a so-called \emph{token state}, which captures the distributional semantics of the code token in its context of use.

After this step, we obtain a set of method feature vectors, one for each method in a file. The next step is aggregating those feature vectors into a single feature vector. The aggregation operation is known as \emph{pooling}.
For example, the simplest \emph{statistical pooling} method is mean-pooling where we take the sum of the method vectors and divide it by the number of methods in a file. More complex pooling methods can be used and we will discuss it in more detail. This step produces a set of \emph{syntactic features} for a file.

Those learned syntactic features are however local to a project. For example, method names and variables are typically project-specific. Hence, using only those features alone may be effective for within-project prediction but may not be sufficient for cross-project settings. Our approach therefore learns another set of features to address this generalization issue. To do so, we build up a universal bag of token states from all files across all the studied projects. We then automatically group those code token states into a number of clusters based on their semantic closeness. The centroids in those clusters form a so-called ``codebook'', which is used for generating a set of \emph{semantic features} for a file through a \emph{semantic pooling} process. We will now describe each of these steps in details.


\subsection{Parsing source code \label{subsec:Parsing-source-code}}

We use Java Abstract Syntax Tree (AST) to extract syntactic information
from source code. To do so, we utilize JavaParser \cite{JavaParser}
to lexically analyze each source file and obtain an AST. Each source
file is parsed into a set of methods and each method is parsed into
a sequence of code tokens. All class attributes (i.e. the header)
are grouped into a sequence of tokens. Comments and blank lines are
ignored. Following standard practice (e.g. as done in \cite{White:2015:TDL}),
we replace integers, real numbers, exponential notation, and hexadecimal
numbers with a generic $\langle num\rangle$ token, and replace constant
strings with a generic $\langle str\rangle$ token. We also replace
less popular tokens (e.g. occurring only once in the corpus) and tokens
which exist in test sets but do not exist in the training set with
a special token $\langle unk\rangle$. A fixed-size vocabulary $\mathscr{V}$
is constructed based on top $N$ popular tokens, and rare tokens are
assigned to $\langle unk\rangle$. Doing this makes our corpus compact
but still provides partial semantic information.



\subsection{Learning code token semantics \label{subsec:token-semantics}}



\begin{figure}[ht]
\centering \includegraphics[width=0.9\linewidth]{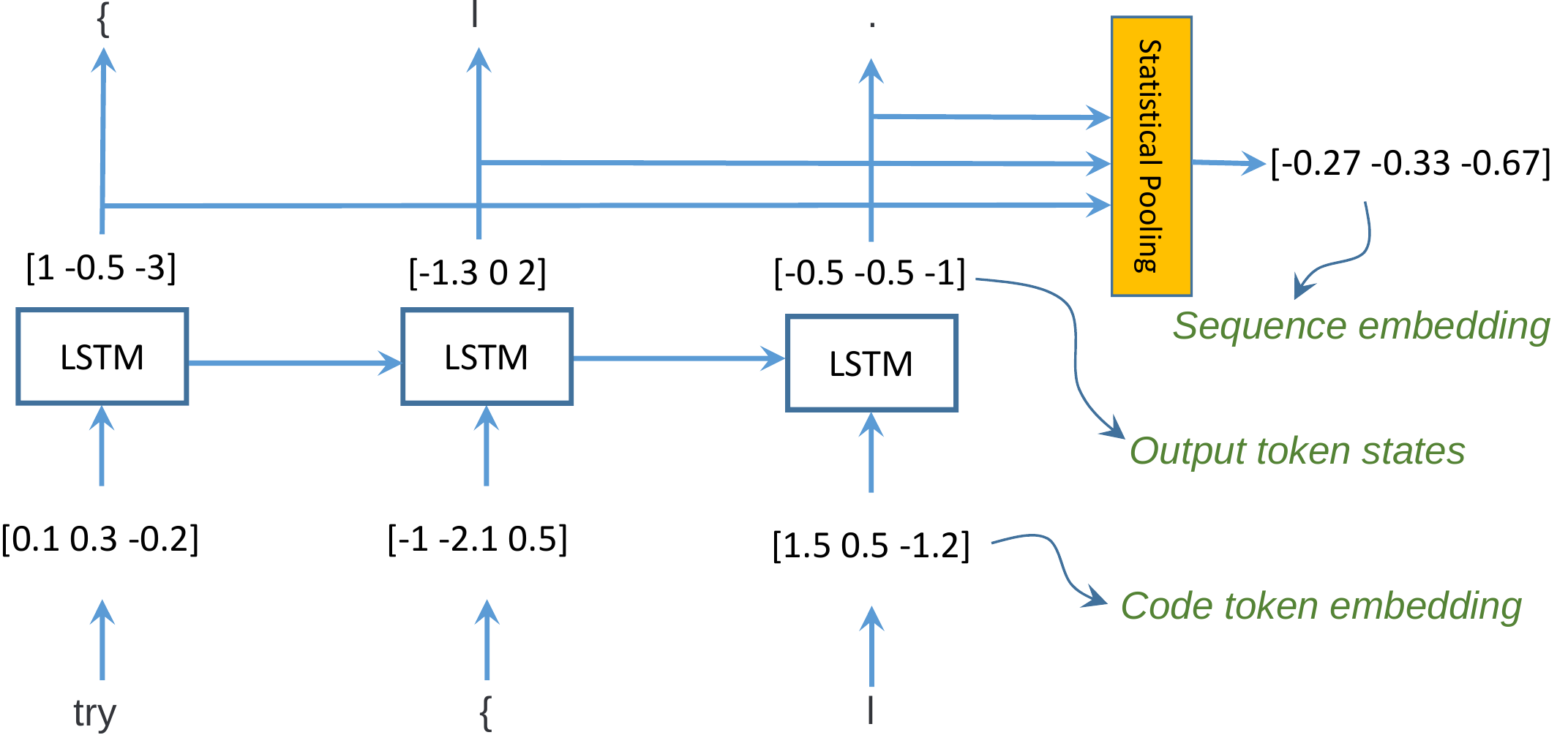}
\caption{An example of how a vector representation is obtained for a code sequence}
\label{fig:LSTM-example-1}
\end{figure}

After the parsing and tokenizing process, each method is now a sequence of code tokens $\langle w_{1},w_{2},...,w_{n}\rangle$. The files in training set give us a large number sequences of code tokens, which are input to an LSTM system. An LSTM unit takes as input a vector representing a code token. Hence, we need to convert each code token into a fixed-length continuous vector. This process is known as \emph{code token embedding}. We do so by maintaining a token embedding matrix $\mathcal{M}\in\mathbb{R}^{d\times|\mathscr{V}|}$ where $d$ is the size of a code token vector and $|\mathscr{V}|$ is the size of vocabulary $\mathscr{V}$. Each code token has an index
in the vocabulary, and this embedding matrix acts as a look-up table:
each column $i^{th}$ in the embedding matrix is an embedded vector
for the token $i^{th}$. We denote $\xb_{t}$ as a vector representation
of code token $w_{t}$. For example, token ``try'' is converted
in vector {[}0.1, 0.3, -0.2{]} in the example in Figure \ref{fig:LSTM-example-1}.

The code sequence vectors that make up each method are then input
to a sequence of LSTM units. Specifically, each token vector $\xb_{t}$
in a sequence $\langle\xb_{1},\xb_{2},...,\xb_{n}\rangle$ is input
into an LSTM unit (see Figure \ref{fig:LSTM-example-1}).
As LSTM is a recurrent net, all the LSTM units share the same model parameters.
Each unit computes the output state $\Sb_{t}$ for an input token $\xb_{t}$.
For example in Figure~\ref{fig:LSTM-example-1}, the output state vector for code token
``try'' is {[}0.1, 0.3, -0.2{]}. The size of this vector can be
different from the size of the input token vector (i.e. $d\neq d'$),
but for simplicity in training the model we assume they are the same.
The state vectors are used to predict the next tokens using another token
weight matrix denoted as $\mathcal{U}\in\mathbb{R}^{d'\times|\mathscr{V}|}$.







LSTM automatically learns both model parameters,  the token weight matrix $\mathcal{U}$ and
the code token embedding matrix $\mathcal{M}$ by maximizing the likelihood
of predicting the next code token in the training data. Specifically, we use the output state vector of code token
$w_{t}$ to predict the next code token $w_{t+1}$ from a context
of earlier code tokens $w_{1:t}$ by computing a posterior distribution:

\begin{equation}
P\left(w_{t+1}=k\mid w_{1:t}\right)=\frac{\exp\left(\mathcal{U}_{k}^{\top}\Sb_{t}\right)}{\sum\limits _{k'}\exp\left(\mathcal{U}_{k'}^{\top}\Sb_{t}\right)}\label{eq:softmax}
\end{equation}
where $k$ is the index of token $w_{t+1}$ in the vocabulary, $\mathcal{U}^{\top}$
is the transpose of matrix $\mathcal{U}$, and $\mathcal{U}_{k}^{\top}$
indicates the vector in column $k^{th}$ of $\mathcal{U}^{\top}$,
and $k'$ runs through all the indices in the vocabulary, i.e. $k'\in\{1,2,...,|\mathscr{V}|\}$.
This learning style essentially estimates a language model of code. 
Thus the LSTM automatically learns a grammar of code \cite{giles2001noisy}.

\subsubsection{Model training}

LSTM can automatically train itself using code sequences in all the methods extracted from our dataset. During training, for every token in a sequence $\langle w_{1},w_{2},...,w_{n}\rangle$, we know the true next token. For example, the true next token after
``try'' is ``\{'' in the example Figure \ref{fig:LSTM-example-1}.
We use this information to learn the model parameters which maximize
the accuracy of our next token predictions. To measure the accuracy,
we use the log-loss (i.e. the cross entropy) of each true next token,
i.e. $-\log P(w_{1})$ for token $w_{1}$, $-\log P(w_{2}\mid w_{1}$)
for token $w_{2}$, ..., $-\log P\left(w_{n}\mid w_{1:n-1}\right)$
for token $w_{n}$. The model is then trained using many known sequences
of code tokens in a dataset by minimizing the following sum log-loss
in each sequence:

\begin{equation}
L(\mathcal{P})=-\log P(w_{1})-\sum_{t=1}^{n-1}\log P\left(w_{t+1}\mid\wb_{1:t}\right)\label{eq:log-loss}
\end{equation}
which is essentially $-\log P\left(w_{1},w_{2},...,w_{n}\right)$.


Learning involves computing the gradient of $L(\mathcal{P})$ during
the back propagation phase, and updating the model parameters $\mathcal{P}$,
which consists of $\mathcal{M}$, $\mathcal{U}$ and other internal LSTM parameters,
via stochastic gradient descent.

\subsubsection{Generating output token states}\label{subsec:token-states}

Once the training phase has been completed we use the learned LSTM to compute a code token state vector $\Sb_{t}$ for every code token $w_{t}$ extracted in our dataset. The use of LSTM ensures that a code token state contains information from other code tokens that come before it. Thus, a code token state captures the \emph{distributional semantics}, a Natural Language Processing concept which dictates that the meaning of a word (code token) is defined by its context of use \cite{baroni2014don}. The same lexical token can theoretically be realized in infinite number of usage contexts. Hence a token semantics is a point in the semantic space defined
by all possible token usages. The token states are then used for generating two distinct sets of features for a file.

\subsection{Generating syntactic features \label{subsec:Method-features}}


Generating syntactic features for a file  involves two steps. First, we generate a set of features for each method in the file. To do so, we first extract a sequence of code tokens $\langle w_{1},w_{2},...,w_{n}\rangle$ from a method, feed it into the trained LSTM system, and obtain an output sequence of token states $\langle\Sb_{1},\Sb_{2},...,\Sb_{n}\rangle$ (see Section \ref{subsec:token-states}). We then compute the method feature vector  by aggregating all the token states in the same sequence so that all information from the start to the end of a method is accumulated (see Figure \ref{fig:LSTM-example-1}). This process is known as \emph{pooling} and there are multiple ways to perform pooling, but the main requirement is that pooling must be length invariant, that is, pooling is not sensitive to variable method lengths. We employ a number of simple but often effective \emph{statistical pooling} methods: (1) Mean pooling, i.e. $\bar{\Sb}=\frac{1}{n}\sum\limits _{t=1}^{n}\Sb_{t}$; (2) Variance pooling, i.e., $\sigmab=\sqrt{\frac{1}{n}\sum\limits _{t=1}^{n}\left(\Sb_{t}-\bar{\Sb}\right)\ast\left(\Sb_{t}-\bar{\Sb}\right)}$, where $\ast$ denotes element-wise multiplication; and (3) A concatenation of both mean pooling and variance pooling, i.e. $\left[\bar{\Sb},\sigmab\right]$.

Since a file contains multiple methods, the next step involves aggregating all these method vectors a single vector for  file. We employ again another statistical pooling mechanism to generate a set of syntactic features for the file.




\subsection{Generating semantic features} \label{subsec:codebook}

Syntactic features are useful for within-project vulnerability prediction since they are local to a method and thus tend to be project-specific. To enable effective cross-project prediction, we need another set of features for a file which reflect how the file positions in a semantic space across all projects. We view a file as a set of code token states (generated from the LSTM system), each of which captures the semantic structure of the token usage contexts. This is different from viewing the file as a Bag-of-Words where a code token is nothing but an index in the vocabulary, regardless of its usage. We partition this set of token states into subsets, each of which corresponds to a distinct region in the semantic space. Suppose there are $k$ regions, each file is then represented as a vector of $k$ dimensions. Each dimension is the number of token state vectors that fall
into the respective region.

The next question is how to partition the semantic space into a number of regions. To do so, we borrow the concept from computer vision by considering each token in a file as an analogy for a salient point (i.e. the most informative point in an image). The token states are akin to the set of point descriptors such as SIFT \cite{lowe1999object}. The main difference here is that in vision, visual descriptors are calculated manually, whereas in our setting token states are learnt automatically through LSTM. In vision, descriptors are clustered into a set of points called \emph{codebook} (not to be confused with the software source code), which is essentially the descriptor centroids.

\begin{figure}[ht]
\centering \includegraphics[width=1\linewidth]{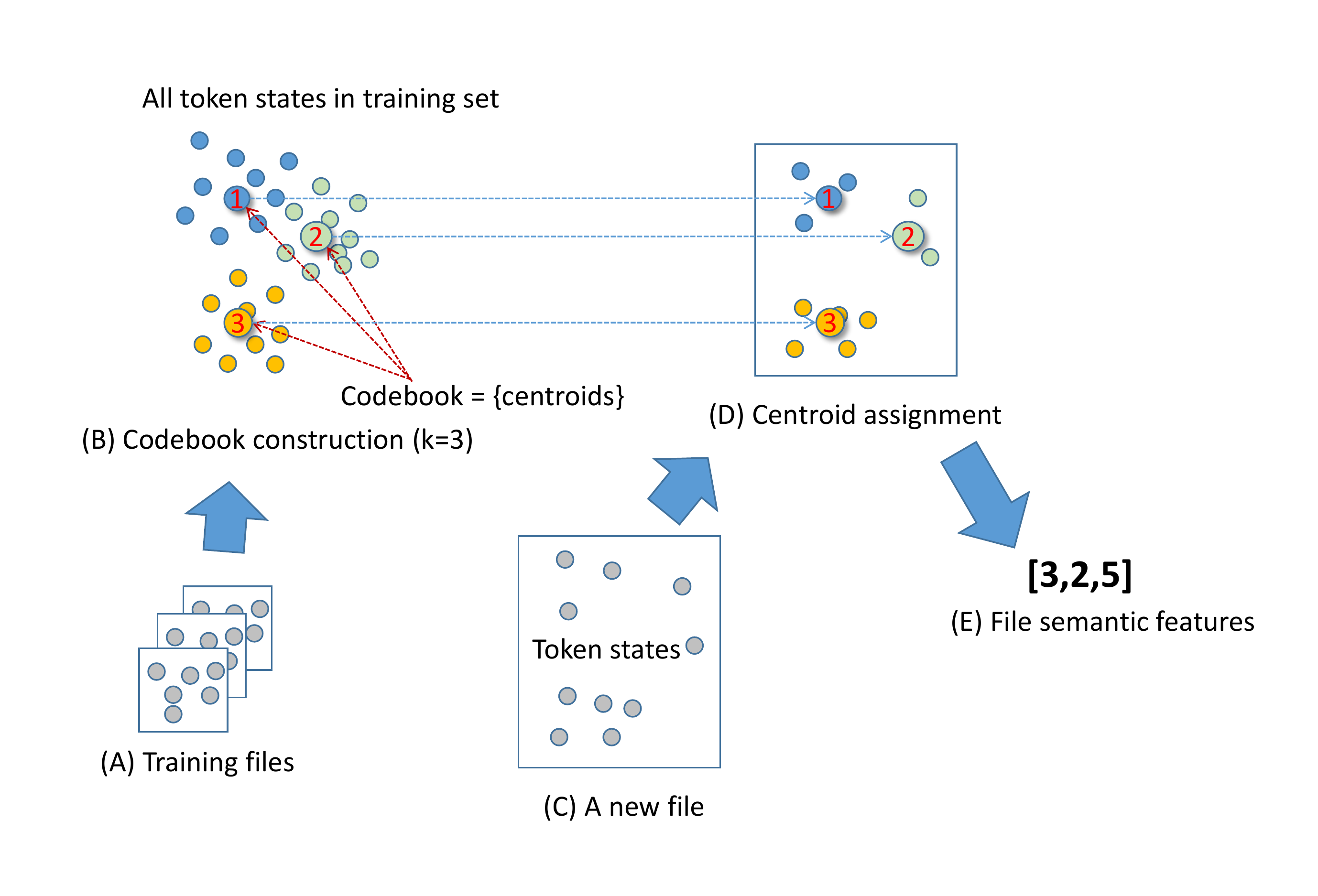}
\caption{An example of using ``codebook'' to automatically learn and generate file semantic features}
\label{fig:codebook}
\end{figure}

Similarly, we can build a codebook that summarizes all token states, i.e. the semantic space, across all projects in our dataset. Each ``code'' in the codebook represents a distinct region in the semantic space. We construct a codebook by using $k$-means to cluster all state vectors in the training set, where $k$ is the pre-defined number of centroids, and hence the size of the codebook (see Part A in Figure \ref{fig:codebook}, each small circle representing a token state in the space). The example in Figure \ref{fig:codebook} uses $k=3$ to produce three state clusters. For each new file, we obtain all the state vectors (Part B in Figure \ref{fig:codebook}) and assign each of them to the closest centroids (Part C in Figure \ref{fig:codebook}). The file is represented as a vector of length $k$, whose elements are the number of centroid occurrences. For example, the new file in Figure \ref{fig:codebook} has 10 token vectors. We then compute the distances between those vectors to the three centroids established in the training set. We find that 3 of them are closest to centroid \#1, 2 to centroid \#2 and 5 to centroid \#3. Hence, the feature vector for the new file is [3, 2, 5].

This technique provides a powerful abstraction over a number of tokens in a file. The intuition here is that the number of tokens in an entire dataset could be large but the number of usage context types (i.e. the token state clusters) can be small. Hence, an file can be characterized by the types of the usage contexts which it contains. This approach offers an efficient and effective way to learn new features for a file from the code tokens constituting it. In effect, a file is a collection of distinct regions in the semantic space of code tokens.






\subsection{Vulnerability prediction}

Using the process above, we generate both syntactic and semantic features for each file in both the training data and the test data. We then use the standard process described in Section \ref{sect:vuln-prediction} to build and train vulnerability prediction models. Files in the test data are used to assess the performance of the trained prediction models.





\section{Implementation}

The proposed approach is implemented in Theano \cite{Theano} and Keras\cite{Keras}
frameworks, running in Python.  Theano supports automatic differentiation of the loss in Eq.~(\ref{eq:log-loss})
and a host of powerful adaptive gradient descent methods. Keras is a wrapper making
model building much easier.

\subsection{Training details}

In particular we use RMSprop as the optimizer with learning rate of 0.02, and smoothing
hyper-parameters: $\rho=0.99$, $\epsilon=1e-7$. The model parameters
are updated in a stochastic fashion, that is, after every mini-batch
of size $50$. We use $|\mathscr{V}|=5,000$ most frequent tokens
for learning the code language model discussed in Section  \label{subsec:token-semantics}. 
We use \emph{dropout} rate of 0.5 at the hidden output of LSTM layer.

The dataset contains more than 240K sequences, in which, 200k sequences are used for training and the others are used for validation. The LSTM which achieved the best perplexity on validation set was finally kept for feature extraction.  The classifier is Random Forest implemented in the scikit-learn toolkit. Hyper-parameters are tuned for best performance include (i) the number of trees, (ii) the maximum depth of a tree, (iii) the minimum number of samples required to split an internal node and (iv) the maximum number of features per tree. The code is run on Intel(R) Xeon(R) CPU E5-2670 0 @ 2.6GHz. There machine has two CPUs, each has 8 physical cores or 16 threads, with a RAM of 128GB.

\subsection{Handling large vocabulary}


To evaluate the prediction probability in Eq.~(\ref{eq:softmax})
we need to iterate through all unique tokens in the vocabulary. Since
the vocabulary's size is large, this can be highly expensive. To tackle the
issue, we employ an approximate method known as Noise Contrastive
Estimation \cite{gutmann2012noise}, which approximates the
vocabulary at each probability evaluation by a small subset of words randomly
sampled from the vocabulary. We use $100$ words, as it is known to work
well in practice \cite{DeepCode2016}.

\subsection{Handling long methods}

Methods are variable in size. This makes learning inefficient because
we need to handle each method separately, not making use of recent
advances in Graphical Processing Units (GPUs). A better way is to
handle methods in mini-batch of fixed size. A typical way is to pad
short methods with dummy tokens so that all methods in the same mini-batch
have the same size. However, since some methods are very long, this
approach will result in a waste of computational effort to handle
dummy tokens. Here we use a simple approach to split a long method
into non-overlapping sequences of fixed length $T$, where $T=100$
is chosen in this implementation due to the faster learning speed.
For simplicity, features of a method are simply the mean features of its sequences.




\section{Evaluation}


\subsection{Datasets}

To carry out our empirical evaluation, we exploited a publicly available dataset \cite{Scandariato2014Dataset} that has been used in previous work \cite{ScandariatoWHJ14} for vulnerability prediction. This dataset originally contained 20 popular applications which were collected from F-Droid and Android OS in 2011. The dataset covers a diversity of application domains such as education, book, finance, email, images and games. However, the provided dataset only contained the application names, their versions (and dates), and the file names and their vulnerability labels. It did not have the source code for the files, which is needed for our study. Using the provided file names and version numbers, we then retrieved the relevant source files from the code repository of each application.

\begin{table}[h]
	\centering
	\caption{Dataset statistics}
	\label{table:dataset-stats}
	\resizebox{3.5in}{!}{%
		\begin{tabular}{@{}lcccccc@{}}
			\toprule
			App         & \#Versions     & \#Files    & Mean files       & Mean LOC   & Mean Vuln   & \% Vuln    \\
			\midrule
            Crosswords & 16 & 842 & 52 &  12,138 & 24 & 0.46 \\
            Contacts&6&787&131&39,492&40&0.31 \\
            Browser&6&433&72&23,615&27&0.37 \\
            Deskclock&6&127&21&4,384&10&0.47 \\
            Calendar&6&307&51&21,605&22&0.44 \\
            AnkiAndroid&6&275&45&21,234&27&0.59 \\
            Mms&6&865&144&35,988&54&0.37 \\
            Boardgamegeek&1&46&46&8,800&11&0.24 \\
            Gallery2&2&545&272&68,445&75&0.28 \\
            Connectbot&2&104&52&14,456&24&0.46 \\
            Quicksearchbox&5&605&121&15,580&26&0.22 \\
            Coolreader&12&423&35&14,708&17&0.49 \\
            Mustard&11&955&86&14,657&41&0.47 \\
            K9&19&2,660&140&50,447&65&0.47 \\
            Camera&6&457&76&16,337&29&0.38 \\
            Fbreader&13&3,450&265&32,545&78&0.30 \\
            Email&6&840&140&51,449&75&0.54 \\
            Keepassdroid&12&1,580&131&14,827&51&0.39 \\
			 \bottomrule
		\end{tabular}
	}
\end{table}

We could not find the code repository for two applications since they appeared no longer available. For some apps, the number of versions we could retrieve from the code repository is less than that in the original datasets. For example, we retrieve the source files for 16 versions of Crossword while the original dataset had 17 versions. The source files for some older versions were no longer maintained in the code repository. Table \ref{table:dataset-stats} provides some descriptive statistics for 18 apps in our dataset, including the number of versions, the total number of files, the average number of files in a version, the average number of lines of code in a version, the average number of vulnerable files in a version, and the ratio of vulnerable files.

\subsection{Research questions}

We followed previous work in vulnerability prediction \cite{ScandariatoWHJ14} and aimed to answer the following standard research questions:

      \begin{enumerate}

        \item \textbf{RQ1. Within-project prediction}: \emph{Are the automatically learned features using LSTM suitable for building a vulnerability prediction model?}

        To answer this question, we focused on one version (the first one) in each application in our dataset. We ran a cross-fold validation experiments by dividing the files in each application into 10 folds, each of which have the approximately same ratio between vulnerable files and clean files. Each fold is used as the test set and the remaining folds are used for training. As a result, we built 10 different prediction models and the performance indicators are averaged out of the 10 folds.

        \item \textbf{RQ2. Cross-version prediction}: \emph{How does our proposed approach perform in predicting future releases, i.e. when the model is trained using an older version in application and tested on a newer version in the same application?}

        In this second experiment, all the files in the first version of each application are used to train a prediction model, which is then used to predict vulnerability of the files of all subsequent versions. For example, the first version in the Crosswords app is used to training and each of the remaining 15 versions is used as a test set.

        \item \textbf{RQ3. Cross-project prediction}:\emph{ Is our approach suitable for cross-project predictions where the model is trained using a source application and tested on a different application?}

            In the third experiment, we used all the files in the first version of each project for training a prediction model. This model is then tested on the first version of the remaining applications.

      \end{enumerate}

\subsection{Benchmarks}

We compare the performance of our approach against the following benchmarks:

\textbf{Software metrics}:  Complexity metrics have been extensively used for defect prediction (e.g. \cite{Hall:2012}) and vulnerability prediction (e.g. \cite{Shin:2011:ECC,Shin:2008:EMP,Chowdhury:2011:UCC}). This is resulted from the intuition that complex code is difficult to understand, maintain and test, and thus has a higher chance of having vulnerabilities than simple code. We have implemented a vulnerability prediction models based on 60 metrics. These features are commonly used in existing vulnerability prediction models. They covers 7 categories:  cohesion metrics, complexity metrics, coupling metrics, documentation metrics, inheritance metrics, code duplication metrics, and size metrics.

\textbf{Bag of Words}: This technique has been used in previous work \cite{ScandariatoWHJ14}, which also consider source code as a special form of text. Hence, it treats a source file as a collection of terms associated with frequencies. The term frequencies are the features which are used as the predictors for a vulnerability prediction model. Lexical analysis is done to source code to break it into a vector of code tokens and the frequency of each token in the file is counted. We also followed previous work \cite{ScandariatoWHJ14} by discretizing the BoW features since they found that this method significantly improved the performance of the vulnerability prediction models. The discretization process involves transforming the numerical BoW features into two \emph{bins}. If a code token occurs more than a certain threshold (e.g. 5 times) than it is mapped to one bin, otherwise it is mapped to another bin.

\textbf{Deep Belief Network}: Recent work \cite{Wang:2016:ALS} has demonstrated that Deep Belief Network (DBN) \cite{hinton2006rdd} worked well for defect prediction. DBN is a family of stochastic deep neural networks which extract multiple layers of data representation. In our implementation, DBN takes the word counts per file as input and produces a latent posterior as output, which is then used as a new file representation. Since the standard DBN accepts only input in the range [0,1], we normalize the word counts by dividing each dimension to its maximum value accross the entire training data. The DBN is then built in a stage-wise fashion as follows. At the bottom layer, a Restricted Boltzmann Machine (RBM) is trained on the normalized word count. An RBM is a special two-layer neural network with binary neurons. 
Following the standard practice in the literature, the RBM is trained using Contrastive Divergence \cite{hinton2006rdd}. After the first RBM is trained, its posterior is used as the input for the next RBM, and the training is repeated. Finally, the two RBMs are stacked on top of each other to form a DBN with two hidden layers. The posterior of the second RBM is used as the new file representation. In our implementation, the two hidden layers have the size of 500 and 128, respectively.

To enable a fair comparison, we used the same classifier for our prediction models and all the benchmarks. We chose Random Forests (RF), an ensemble method which combines the estimates from multiple estimators since it is one of the most effective classifier for vulnerability prediction \cite{ScandariatoWHJ14}. We employed the standard Precision, Recall, and F-measure, which has been widely-used in the literature (e.g. \cite{ScandariatoWHJ14,Shin:2011:ECC,Chowdhury:2011:UCC,Zimmermann:2010:SNH}) for evaluating the predictive performance of vulnerability prediction models built using the three above benchmarks and our approach.

\subsection{Results}

\begin{table*}[h]
\centering \caption{Within-project results (RQ1) for the three benchmarks and three variations of our approach. ``Joint features'' indicates the use of both syntactic features and semantic features.}
\label{table:results-RQ1} \resizebox{7.0in}{!}{%
\begin{tabular}{@{}l|ccc|ccc|ccc|ccc|ccc|ccc}
\toprule
\multirow{2}{*}{Application} & \multicolumn{3}{c}{Software metrics} & \multicolumn{3}{c}{Bag-of-Words} & \multicolumn{3}{c}{Deep Belief Network} &  \multicolumn{3}{c}{\textbf{Syntactic features}} & \multicolumn{3}{c}{\textbf{Semantic features}} & \multicolumn{3}{c}{\textbf{Joint features}}\tabularnewline
 & {}P {} & {}R{} & {}F{} & {}P{} & {}R{} & {}F{} & {}P{} & {}R{} & {}F{} &  {}P{} & {}R{} & {}F{} & {}P{} & {}R{} & F{} & {}P{} & {}R{} & {}F{}\tabularnewline
\midrule
AnkiAndroid  & 0.62 & 0.65 & 0.61       & 1.00 & 0.94 & 0.96        & 1.00 & 0.94 & 0.96        & 1.00 & 1.00 & 1.00 & 1.00 & 0.94  & 0.96 & 1.00 & 0.94 & 0.96\tabularnewline
Boardgamegeek  & 0.18 & 0.35 & 0.23     & 0.95 & 1.00 & 0.97      & 0.95 & 1.00 & 0.97      & 0.95 & 0.95 & 0.93 & 0.90 & 0.90 & 0.90 & 0.90 & 0.90 & 0.90\tabularnewline
Browser  & 0.65 & 0.60 & 0.62 &         0.96 & 1.00 & 0.98          & 0.96 & 1.00 & 0.98         & 1.00  & 1.00 & 1.00 & 0.96 & 1.00 & 0.98 & 0.96 & 1.00 & 0.98\tabularnewline
Calendar  & 0.52 & 0.43 & 0.45              & 0.89 & 0.80 & 0.83        & 1.00 & 0.91 & 0.94        & 1.00  & 1.00 & 1.00 & 0.96 & 0.94 & 0.94 & 1.00 & 1.00 & 1.00\tabularnewline
Camera  & 0.72 & 0.81 & 0.73            & 0.82 & 0.92 & 0.86        & 0.83 & 0.93 & 0.87            & 0.88  & 1.00 & 0.93 & 0.92 & 0.98 & 0.94 & 0.92 & 0.98 & 0.94\tabularnewline
Connectbot  & 0.65 & 0.73 & 0.64    & 0.97 & 0.90 & 0.91        & 0.92 & 0.90 & 0.88            & 1.00 & 0.97 & 0.98 & 0.93 & 0.93 & 0.93 & 0.97 & 0.93 & 0.93\tabularnewline
Contacts  & 0.72 & 0.72 & 0.71          & 0.94 & 0.88 & 0.89        & 0.92 & 0.81 & 0.83             & 0.96  & 0.94 & 0.95 & 0.89  & 0.94 & 0.89 & 0.87 & 0.97 & 0.90\tabularnewline
Coolreader  & 0.55 & 0.70 & 0.60        & 1.00 & 1.00 & 1.00        & 1.00 & 1.00 & 1.00            & 1.00 & 0.94 & 0.96 & 1.00 & 0.94 & 0.96 & 1.00 & 0.94 & 0.96\tabularnewline
Crosswords  & 0.55 & 0.57 & 0.55        & 0.89 & 0.89 & 0.89    & 0.89 & 0.89 & 0.89         & 0.89 & 0.89 & 0.89 & 0.89 & 0.83 & 0.85 & 0.89 & 0.83 & 0.85\tabularnewline
Deskclock  & 0.30 & 0.20 & 0.23         & 0.92 & 0.92 & 0.89    & 0.83 & 0.92 & 0.83     & 1.00 & 1.00  & 1.00 & 1.00 & 0.92 & 0.94 & 1.00 & 0.92 & 0.94\tabularnewline
Email  & 0.83 & 0.84 & 0.82         & 0.92 & 0.93 & 0.92        & 0.96 & 0.94 & 0.94        & 0.93 & 0.95 & 0.93 & 0.95 & 0.90 & 0.92 & 0.94 & 0.93 & 0.93\tabularnewline
Fbreader  & 0.60 & 0.52 & 0.55      & 0.78 & 0.88 & 0.81        & 0.81 & 0.85 & 0.82        & 0.83 & 0.82 & 0.82 & 0.85 & 0.83 & 0.84 & 0.82 & 0.86 & 0.83\tabularnewline
Gallery2  & 0.62 & 0.60 & 0.57      & 0.80 & 0.91 & 0.84        & 0.75 & 0.87 & 0.78        & 0.83 & 0.88 & 0.84 & 0.83 & 0.87 & 0.84 & 0.80 & 0.89 & 0.83\tabularnewline
K9  & 0.73 & 0.82 & 0.76        & 0.94 & 0.88 & 0.89        & 0.96 & 0.87 & 0.90        & 0.89 & 1.00 & 0.94 & 0.88 & 0.97 & 0.91 & 0.88 & 0.98 & 0.92\tabularnewline
Keepassdroid  & 0.69 & 0.75 & 0.71      & 0.91 & 0.88 & 0.88        & 0.85 & 0.90 & 0.87        & 0.93 & 0.97 & 0.94 & 0.90 & 0.91 & 0.89 & 0.88 & 0.89 & 0.88\tabularnewline
Mms  & 0.78 & 0.60 & 0.67       & 0.86 & 0.85 & 0.85        & 0.84 & 0.85 & 0.84        & 0.87 & 0.85 & 0.85 & 0.87 & 0.88 & 0.87 & 0.87 & 0.92 & 0.89\tabularnewline
Mustard  & 0.81 & 0.68 & 0.72       & 0.94 & 0.93 & 0.93        & 0.94 & 0.87 & 0.89        & 0.96 & 0.99 & 0.97 & 0.97 & 0.93 & 0.94 & 0.95 & 0.99 & 0.96\tabularnewline
Quicksearchbox  & 0.60 & 0.42 & 0.43    & 0.84 & 0.90 & 0.83        & 0.83 & 0.90 & 0.84        & 0.93 & 0.90 & 0.88 & 0.89 & 0.77 & 0.82 & 0.89 & 0.79 & 0.82\tabularnewline
\midrule
Average  & 0.62 & 0.61 & 0.59       & 0.91 & 0.91 & 0.90         & 0.90 & 0.91 & 0.89       & \textbf{0.94} & \textbf{0.95} & \textbf{0.93} & \textbf{0.92} & \textbf{ 0.91} & \textbf{0.91} & \textbf{0.92} & \textbf{ 0.93} & \textbf{0.91}\tabularnewline
\bottomrule
\end{tabular}}
\end{table*}

\subsubsection{Learned code token semantics}

\begin{figure}[ht]
	\centering
	\includegraphics[width=\linewidth]{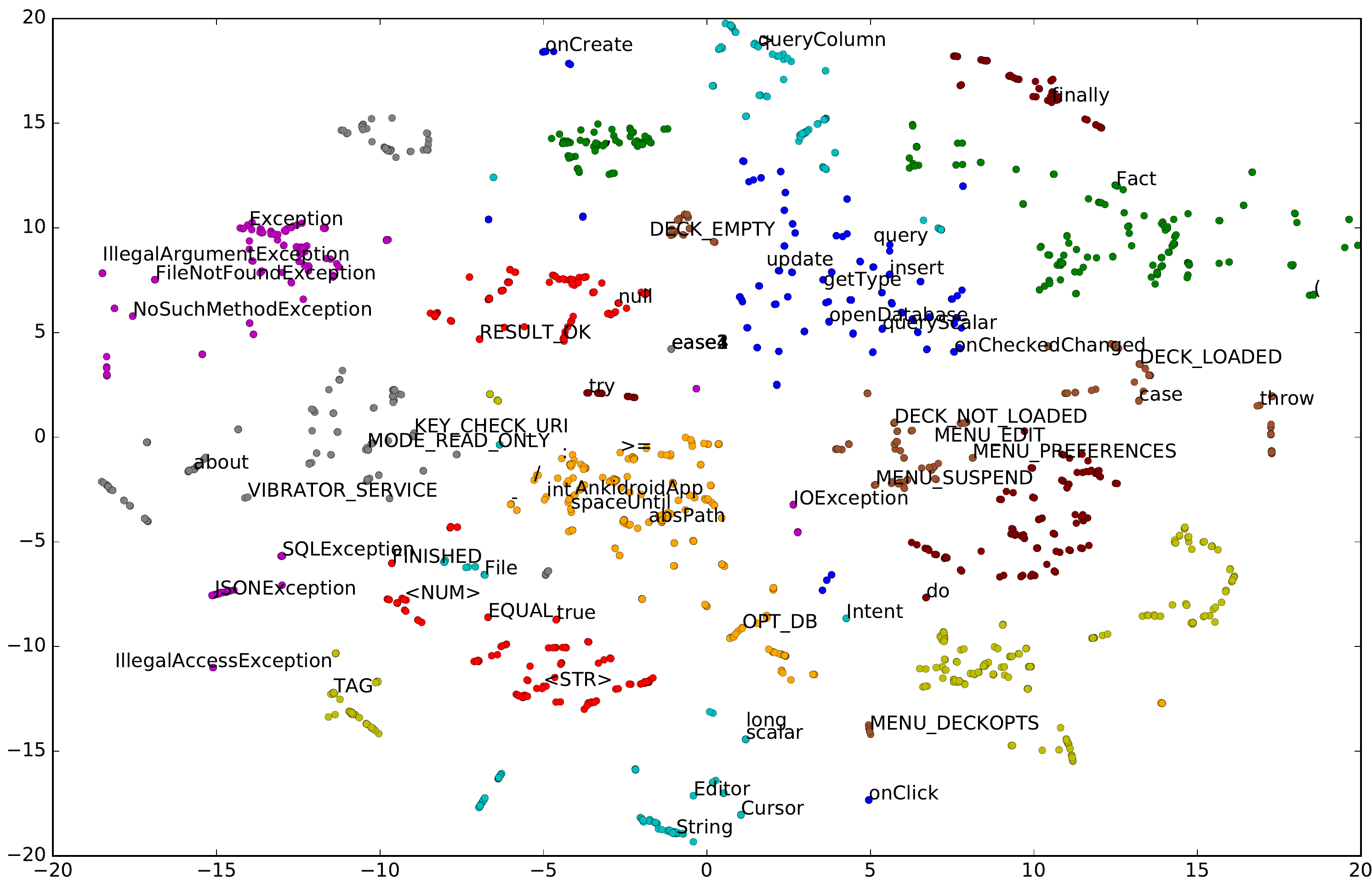}
	\caption{Top 2,000 frequent code tokens were automatically grouped into clusters (each cluster has a distinct color)}
	\label{figure:code-clusters}
\end{figure}

An important part of our approach is learning the semantics of code tokens using the context of its usage through LSTM. Figure \ref{figure:code-clusters} show the top 2,000 frequent code tokens used in our dataset. They were automatically grouped in 10 clusters (using K-means clustering) based on their token states learned through LSTM. Recall that these clusters are the basis for us to construct a codebook (discussed in Section \ref{subsec:codebook}). We used t-distributed stochastic neighbor embedding (t-SNE) \cite{van2008visualizing} to display high-dimensional vectors in two dimensions. We show here some representative code tokens from some clusters for a brief illustration. Code tokens that are semantically related are grouped in the same cluster. For example, code tokens related to exceptions such as IllegalArgumentException, FileNotFoundException, and NoSuchMethodException are grouped in one cluster. This indicates, to some extent, that the learned token states effectively capture the semantic relations between code tokens, which is useful for us to learn both syntactic and semantic features later.


\begin{table*}[h]
\centering \caption{Cross-version results (RQ2) for the three benchmarks and three variations of our approach. ``Joint features'' indicates the use of both syntactic features and semantic features.}
\label{table:results-RQ2} \resizebox{7.0in}{!}{%
\begin{tabular}{@{}l|ccc|ccc|ccc|ccc|ccc|ccc}
\toprule
\multirow{2}{*}{Application} & \multicolumn{3}{c}{Software metrics} & \multicolumn{3}{c}{Bag-of-Words} & \multicolumn{3}{c}{Deep Belief Network} & \multicolumn{3}{c}{\textbf{Syntactic features}} & \multicolumn{3}{c}{\textbf{Semantic features}} & \multicolumn{3}{c}{\textbf{Joint features}}\tabularnewline
 &  {}P{} & {}R{} & {}F{} & {}P{} & {}R{} & {}F{} & {}P{} & {}R{} & {}F{} & {}P{} & {}R{} & {}F{} & {}P{} & {}R{} & F{} & {}P{} & {}R{} & {}F{}\tabularnewline
\midrule
AnkiAndroid  & 0.80 & 0.71 & 0.75   & 0.88 & 0.85 & 0.87    & 0.87 & 0.85 & 0.86     & 0.83 & 0.90 & 0.87 & 0.91 & 0.84 & 0.87 & 0.89 & 0.87 & 0.88\tabularnewline
Browser  & 0.69 & 0.65 & 0.66   & 0.75 & 0.63 & 0.68    & 0.75 & 0.65 & 0.70     & 0.59 & 0.64 & 0.61 & 0.84 & 0.57 & 0.68 & 0.82 & 0.58 & 0.68\tabularnewline
Calendar  & 0.69 & 0.77 & 0.72      & 0.74 & 0.83 & 0.78    & 0.71 & 0.85 & 0.78    & 0.73 & 0.81 & 0.77 & 0.79 & 0.89 & 0.83 & 0.79 & 0.84 & 0.82\tabularnewline
Camera  & 0.51 & 0.81 & 0.62    & 0.60 & 0.90 & 0.72    & 0.58 & 0.89 & 0.70    & 0.65 & 0.84 & 0.73 & 0.67 & 0.88 & 0.76 & 0.84 & 0.80 & 0.82\tabularnewline
Connectbot  & 0.75 & 0.75 & 0.75    & 1.00 & 0.96 & 0.98    & 1.00 & 0.96 & 0.98    & 1.00 & 1.00 & 1.00 & 1.00 & 1.00 & 1.00 & 1.00 & 1.00 & 1.00\tabularnewline
Contacts  & 0.68 & 0.72 & 0.70      & 0.74 & 0.71 & 0.73    & 0.73 & 0.75 & 0.74    & 0.60 & 0.80 & 0.69 & 0.69 & 0.72 & 0.70 & 0.73 & 0.69 & 0.71\tabularnewline
Coolreader  & 0.80 & 0.77 & 0.78    & 0.93 & 0.79 & 0.85     & 0.83 & 0.82 & 0.83    & 0.94 & 0.89 & 0.91 & 0.89 & 0.85 & 0.87 & 0.92 & 0.86 & 0.89\tabularnewline
Crosswords  & 0.87 & 0.72 & 0.79    & 0.90 & 0.81 & 0.86    & 0.83 & 0.88 & 0.85     & 0.85 & 0.89 & 0.87 & 0.88 & 0.90 & 0.89 & 0.92 & 0.83 & 0.87\tabularnewline
Deskclock  & 0.87 & 0.66 & 0.75     & 0.89 & 0.80 & 0.85    & 0.87 & 0.76  & 0.81    & 0.93 & 0.80 & 0.86 & 0.90 & 0.86 & 0.88 & 0.89 & 0.82 & 0.86\tabularnewline
Email       & 0.74 & 0.83 & 0.78    & 0.82 & 0.83 & 0.82    & 0.79 & 0.85 & 0.82     & 0.77 & 0.93 & 0.84 & 0.86 & 0.89 & 0.87 & 0.83 & 0.90 & 0.86\tabularnewline
Fbreader  & 0.83 & 0.75 & 0.78      & 0.79 & 0.83 & 0.81    & 0.80 & 0.75 & 0.77     & 0.84 & 0.81 & 0.83 & 0.91 & 0.84 & 0.88 & 0.86 & 0.84 & 0.85\tabularnewline
Gallery2  & 0.74 & 0.74 & 0.74      & 0.96 & 0.93 & 0.95        & 0.88 & 0.88 & 0.88        & 0.95 & 0.97 & 0.96 & 0.96 & 0.99 & 0.97 & 0.96 & 0.97 & 0.97\tabularnewline
K9  & 0.77 & 0.89 & 0.82    & 0.85 & 0.84 & 0.84    & 0.84 & 0.86 & 0.85    & 0.80 & 0.94 & 0.87 & 0.84 & 0.98 & 0.90 & 0.89 & 0.92 & 0.90\tabularnewline
Keepassdroid  & 0.92 & 0.91 & 0.91      & 0.92 & 0.87 & 0.89        & 0.86 & 0.86 & 0.86    & 1.00  & 0.99 & 0.99 & 0.98 & 0.91 & 0.94 & 0.99 & 0.92 & 0.95\tabularnewline
Mms  &  0.83 & 0.81 & 0.82      & 0.93 & 0.90 & 0.91        & 0.94 & 0.86 & 0.90    & 0.93 & 0.89 & 0.91 & 0.93 & 0.93 & 0.93 & 0.92 & 0.96 & 0.94\tabularnewline
Mustard  & 0.93 & 0.90 & 0.91   & 0.97 & 0.96 & 0.96    & 0.92 & 0.93 & 0.93    & 0.98 & 0.98 & 0.98 & 0.99 & 0.95 & 0.97 & 0.99 & 0.95 & 0.97\tabularnewline
Quicksearchbox  & 0.57 & 0.69 & 0.62    & 0.71 & 0.73 & 0.72    & 0.56 & 0.79 & 0.66    & 0.76 & 0.80 & 0.78 & 0.74 & 0.87 & 0.80 & 0.77 & 0.88 & 0.82\tabularnewline
\midrule
Average  & 0.76 & 0.77 & 0.76   & 0.85 & 0.83 & 0.84    & 0.81 &  0.84 & \textbf{ }0.82     & 0.83 & \textbf{0.87} & \textbf{0.85} & \textbf{0.87} &  \textbf{0.87} & \textbf{0.87} & \textbf{0.88} & \textbf{0.86} & \textbf{0.87}\tabularnewline
\bottomrule
\end{tabular}}
\end{table*}

\subsubsection{Within-project prediction (RQ1)}

We experimented with a number of variations of our approach by varying the pooling techniques (mean pooling and standard deviation pooling) and the use of syntactic and semantic features. Table \ref{table:results-RQ1} reports the precision, recall and F-measure for the three benchmarks (software metrics, Bag-of-Words, and Deep Belief Network) and three variations of our approach: using only syntactic features, using only semantic features, and using a joint set of both features types. Note that both the syntactic feature option and  and the joint feature option we reported here used mean pooling for generating method features and standard deviation pooling for generating syntactic file features. The full report which contains the results for all variations of pooling is available at \url{https://goo.gl/OnRnPp}.

Our approach obtained a strong result: in \emph{all} 18 applications, they all achieved more than 80\% in precision, recall, and F-measure (see Table \ref{table:results-RQ1}). Among the three variation of our approach, using only syntactic features appeared to be the best option for vulnerable prediction within one version of a project. This option delivered the highest precision, recall and F-measure averaging across 18 applications. This result is consistent with our underlying theory that syntactic features are project-specific and are thus useful for within-project prediction. Our approach outperforms the three benchmarks to varying extents. The syntactic feature option offered 58\% F-measure improvement over the software metrics approach. Both Bag-of-Words (BoW) and Deep Belief Network approaches also achieved very good results (on average 89--90\% F-measure) -- which is consistent with the findings in previous work \cite{ScandariatoWHJ14} in case of BoW. Hence, comparing against these two benchmarks, the improvement brought by our approach is small (approximately 3\% in F-measure).

\subsubsection{Cross-version prediction (RQ2)}

Table \ref{table:results-RQ2} reports the results in a cross-version setting where a prediction model was trained using the first version of an application and tested using the subsequent versions of the same applications. Our approach again slightly outperforms the BoW and DBN benchmarks. Among the three variations, using semantic features (alone or joint with syntactic features) becomes the best options. This is because semantic measures improve the generalization of the prediction models, which becomes slightly more useful in cross-version settings.

\subsubsection{Cross-project prediction (RQ3)}

This experiment followed the setup in previous work \cite{ScandariatoWHJ14}. We first built 18 prediction models, each of which use the first version of each application for training. Each model was then tested using the first version of the other 17 applications. We ran this experiment with the three benchmarks and variations of our approach. We used the same baseline as in previous work \cite{ScandariatoWHJ14}: a model is applicable to a tested application if both precision and recall are above 80\%. For each application, Table \ref{table:results-RQ3} reports the number of other applications to which the corresponding models can be applied.

\begin{table}[h]
\centering \caption{Cross-project results (RQ3) for the three benchmarks and three variations of our approach. }
\label{table:results-RQ3} \resizebox{3.4in}{!}{%
\begin{tabular}{lcccccc}
\toprule
App & Metrics  & BoW & DBN & \textbf{Syntactic} & \textbf{Semantic} & \textbf{Joint}\tabularnewline
\midrule
AnkiAndroid  & 1  & 2 & 2   & 3 & 7 & 5\tabularnewline
Boardgamegeek  & 0  & 1 & 1     & 0 & 1 & 1\tabularnewline
Browser  & 0  & 1 & 2       & 0 & 0 & 1\tabularnewline
Calendar  & 0  & 1 & 1      & 1 & 4 & 5\tabularnewline
Camera  & 1  & 3 & 2     & 3 & 6 & 6\tabularnewline
Connectbot  & 0  & 2 & 2    & 4 & 4 & 5\tabularnewline
Contacts  & 0  & 1 & 3      & 1 & 2 & 2\tabularnewline
Coolreader  & 1  & 2 & 3    & 4 & 4 & 6\tabularnewline
Crosswords  & 0  & 3 & 3    & 2 & 1 & 1\tabularnewline
Deskclock  & 0  & 1 & 1     & 0 & 1 & 1\tabularnewline
Email  & 1  & 2 & 2     & 4 & 4 & 4\tabularnewline
Fbreader  & 0  & 2 & 2      & 4 & 6 & 6\tabularnewline
Gallery2  & 0  & 3 & 2       & 3 & 3 & 2\tabularnewline
K9  & 1  & 3 & 3        & 2 & 7 & 8\tabularnewline
Keepassdroid  & 0  & 3 & 1      & 4 & 2 & 4\tabularnewline
Mms  & 0  & 4 & 2       & 3 & 5 & 6\tabularnewline
Mustard  & 0  & 3 & 3       & 3 & 8 & 8\tabularnewline
Quicksearchbox  & 0  & 2 & 2        & 1 & 3 & 3\tabularnewline
\midrule
Average & 0.3 & 2.2 & 2.1    & \textbf{2.3} & \textbf{3.8} & \textbf{4.1}\tabularnewline
\bottomrule
\end{tabular}}
\end{table}

The results suggest that using semantic features really improve the general applicability of prediction models. All the models using both semantic and syntactic features were successfully applicable to at least one other application. Some of them (e.g. K9 and Mustard) are even applicable to 8 other applications. In this cross-project prediction setting, our approach also offers bigger improvements over the BoW and DBN benchmarks. On average, a joint-feature model is applicable to around 4 other applications, which is 85\% improvement compared with BoW or DBN models.

\subsection{Remarks and implications}

The high performance of BoW on within-project prediction (RQ1 and RQ2) is not totally surprising for two reasons. One is that BoW has been known as a strong representation for text classification, and source code is also a type of text written in programming languages. The other reason is that although the train files and test files are not identical, a project typically has many versions, and the new versions of the same file may carry a significant amount of information from the old versions. The repeated information can come from multiple forms: fully repeated pieces of code, the same BoW statsistics, or the same code convention and style. Thus any representation that is sufficiently expressive and coupled with highly flexible classifiers such as Random Forests, will likely to work well.

However, this is not the case for cross-project prediction (RQ3). This is when the BoW statistics are likely to be different between projects, and knowledge learned from one project may not transfer well to others. In machine learning and data mining, this problem is known as \emph{domain adaptation}, where each project is a domain. The common approach is to learn the common representation across domains, upon which classifiers will be built. This is precisely what is done using the LSTM-based language model and codebook construction. Note that the LSTM and codebook are learned using all available data without supervision. This suggests that we can actually use external data, even if there are no vulnerability labels. The competitive performance of the proposed deep learning approach clearly demonstrates the effectiveness of this representation learning.

To conclude, when doing within-project prediction, it is useful to use BoW due to its simplicity. But when generalizing from one project to another, it is better to use representation learning. We recommend using LSTM for language model, and codebook for semantic structure discovery.

\section{Threats to validity}

There are a number of threats to the validity of our study, which we discuss below.

\textbf{Construct validity:} We mitigated the construct validity concerns by using a publicly available dataset that has been used in previous work \cite{ScandariatoWHJ14}. The dataset contains real Android applications and vulnerability labels of the files in those applications. The original dataset did not unfortunately contain the source files. However, we have carefully used the information (e.g. application details, version numbers and date) provided with the dataset to retrieve the relevant source files from the code repository of those applications.

\textbf{Conclusion validity:} We tried to minimize threats to conclusion validity by using standard performance measures for vulnerability prediction \cite{ScandariatoWHJ14,Walden:2014:PVC,Shin:2011:ECC,Chowdhury:2011:UCC}. We however acknowledge that a number of statistical tests \cite{STVR:STVR1486} can be applied to verify the statistical significance of our conclusions. Although we have not seen those statistical tests being used in previous work in vulnerability prediction, we plan to do this investigation in our future work.

\textbf{Internal validity:} The dataset we used contains vulnerability labels only for Java source files. In practice, other files (e.g. XML manifest files) may contain security information such as access rights. Another threat concerns the cross-version prediction where we replicated the experiment done in \cite{ScandariatoWHJ14} and allowed that the exactly same files might be present between versions. This might have inflated the results, but all the prediction models which we compared against in our experiment benefit from this.

\textbf{External validity:} We have considered a large number of applications which differ significantly in size, complexity, domain, popularity and revision history. We however acknowledge that our data set may not be representative of all kinds of Android applications. Further investigation to confirm our findings for other Android applications as well as other types of applications such as web applications and applications written in other programming languages such as PhP and C++.

\section{Related work}

Machine learning techniques have been widely used to build vulnerability prediction models. Early approaches (e.g. \cite{Shin:2008:EMP}) employed complexity metrics such as (e.g. McCabe's  cyclomatic complexity, nesting complexity, and size) as the predictors. Later approaches enriched this software metric feature set with coupling and cohesion metrics (e.g. \cite{Chowdhury:2011:UCC}), code churn and developer activity metrics (e.g. \cite{Shin:2011:ECC})), and dependencies and organizational measures (e.g. \cite{Zimmermann:2010:SNH}). Those approaches require knowledgeable domain experts to determine the metrics that are used as predictors for vulnerability.

Recent approaches treat source code as another form of text and leverage text mining techniques to extract the features for building vulnerability prediction models. The work in \cite{ScandariatoWHJ14} used the Bag-of-Words representation in which a source code file is viewed as a set of terms with associated frequencies. They then used the term-frequencies as the features for predicting vulnerability. The BoW approach eliminates the need for manually designing the features. BoW models also produced higher recall than software metric models for PHP applications \cite{Walden:2014:PVC}. However, the BoW approaches carries the inherent limitation of BoW in which syntactic information such as code order is disregarded.


Predicting vulnerabilities is related to software defect prediction. The study in \cite{ShinW13} found that some defect prediction models can be adapted for vulnerability prediction. While code metrics were commonly used as features for building defect prediction models \cite{Hall:2012}, various other metrics have also been employed such as change-related metrics \cite{Moser:2008,Nagappan:2005:URC}, developer-related metrics \cite{Pinzger:2008:DNP}, organization metrics \cite{Nagappan:2008:IOS}, and change process metrics \cite{Hassan:2009:PFU}. Recently, a number of approaches (e.g. \cite{Yang:2015:DLJ,Wang:2016:ALS}) have leveraged a deep learning model called Deep Belief Network \cite{DBN-Hinton06} to automatically learn features for defect prediction.

Deep learning has recently attracted increasing interests in software engineering. The work in \cite{DeepSoftFSE2016} proposes generic deep learning framework based on LSTM for modeling software and its development process. The work in \cite{White:2015:TDL} demonstrated the effectiveness of using recurrent neural networks (RNN) to model source code. Their later work \cite{White:2016:DLC} extended these RNN models for detecting code clones. The work in \cite{Gu:2016:DAL} uses a special RNN  Encoder--Decoder, which consists of an encoder RNN to process the input sequence and a  decoder  RNN  with  attention  to  generate  the  output  sequence, to generate  API  usage  sequences  for a  given  API-related  natural  language  query. The work in \cite{DBLP:conf/aaai/GuptaPKS17} also uses RNN Encoder--Decoder but for fixing common errors in C programs. The work \cite{DeepCode2016} showed that LSTM is even more effective in code modeling, which inspired us to use it for learning vulnerability features. The work in \cite{Huo:2016:LUF} uses Convolutional Neural Networks (CNN) \cite{Cun:1990:HDR} for bug localization.

\section{Conclusions and future work}

This paper proposes to leverage Long-Short Term Memory, a representation deep learning model, to automatically learn features directly from source code for vulnerability prediction. The learned syntactic features capture the sequential structure in code at the method level, while semantic features characterize a source code file by usage contexts of its code tokens. We performed an evaluation on 18 Android applications from a public dataset provided in previous work \cite{ScandariatoWHJ14}. The results for within-project prediction  demonstrate that the automatically learned features significantly outperforms the traditional software metrics approach (58\% improvement on average), and offers a small improvement (3\% on average) over the Bag-of-Word approach and another deep learning approach (Deep Belief Network). For cross-project prediction, the results suggest that our approach is clearly superior to these state-of-the-art techniques (85\% improvement on average).

Our future work involves applying the proposed approach to other types of applications (e.g. Web applications) and programming languages (e.g. PHP or C++) where vulnerability datasets are available. We also aim to leverage our approach to learn features for predicting vulnerabilities at the method and code change levels. In addition, we plan to explore how our approach can be extended to predicting general defects and safety-critical hazards in code. Finally, our future investigation involves building a fully end-to-end prediction system from raw input data (code tokens) to vulnerability outcomes.

\section*{Acknowledgement}

The paper is supported by Samsung 2016 Global Research Outreach Program (GRO) entitled ``Predicting hazardous software components using deep learning''.

\IEEEpeerreviewmaketitle


\bibliographystyle{IEEEtran}
\bibliography{vision,hoa_ref,bib_ref,lstm_ref,security}
\end{document}